# Fracture Properties of Green Nano Fibrous Network with Random and Aligned Fibre Distribution: A Hierarchical Molecular Dynamics and Peridynamics Approach

Razie Izadi*, Raj Das, Nicholas Fantuzzi, Patrizia Trovalusci

*Corresponding author's e-mail: razie.izadi@uniroma1.it

**Abstract:** Polylactic acid (PLA) nanofibrous networks have gained substantial interest across various engineering and scientific disciplines, such as tissue engineering, drug delivery, and filtration, due to their unique and multifunctional attributes, including biodegradability, tuneable mechanical properties, and surface functionality. However, predicting their mechanical behaviour remains challenging due to their structural complexity, multiscale features, and variability in material properties.

This study presents a hierarchical approach to investigate the fracture phenomena in both aligned and randomly oriented nanofibrous networks by integrating atomistic modelling and non-local continuum mechanics, peridynamics. At the nanoscale, all-atom molecular dynamics simulations are employed to apply tensile loads to freestanding pristine and silver-doped PLA nanofibres, where key mechanical properties such as Young's modulus, Poisson's ratio, and critical energy release rate are determined using innovative approaches. A new method is introduced to seamlessly transfer data from molecular dynamics to peridynamics by ensuring the convergence of the tensile response of a single fiber in both frameworks. This nano to micro coupling technique is then utilised to examine the Young's modulus, fracture toughness of mode I and II, and crack propagation in PLA nanofibrous networks. The proposed framework can also incorporate the effects of surface coating and fiber arrangements on the measured properties. The current research paves the way for the development of stronger and more durable eco-friendly nanofibrous networks with optimised performance.

1. Introduction

Polylactic acid nanofibrous networks possess unique and multifunctional features making them ideal for a variety of science and engineering applications. The biodegradability and biocompatibility of PLA nanofibrous network along with tuneable mechanical properties and surface functionality, have generated extensive interest in their application in biomedical sectors, such as tissue engineering [1-6], drug delivery [7] and wound dressing [8] where they primarily serve as mechanical barriers while additionally facilitating cell growth and controlling the release of therapeutic agents. Beyond the biomedical field, the high surface-volume ratio and porosity of PLA nanofibrous networks enhance their ability to immobilise particles and contaminants, suitable for air filtration, adsorption, and water treatment [9]. The high surface area and flexibility also provide PLA nanofibrous networks a notable capacity for sensor application, whilst their electrochemical features further broaden their utility in batteries and super capacitors [10].

However, pristine PLA nanofibres are limited in their functionalities due to their suboptimal mechanical and chemical properties, such as inherent brittleness, low fracture toughness [11] and lack of antibacterial features [12, 13]. Different techniques have been adopted to modify the chemical composition and surface morphology of nanofibrous networks including coating and decorating with metallic nanoparticles, biomolecules, enzymes, carbon nanotubes, surfactants, and polymers [14]. To enhance PLA's functionality, various nanomaterials, including carbon nanotubes (CNTs), graphene, nano clays, natural compounds, nanoceramics, and metallic and metallic oxide nanoparticles, are employed as reinforcements [15]. Each nano additive endows distinct properties to the resulting nanocomposites, though they also introduce certain challenges. For instance, carbon-based nanofillers such as carbon nanotubes, carbon nanofibers, and graphene significantly improve the mechanical properties of PLA, provided they are well dispersed within the polymer matrix. However, significant concerns regarding the toxicity of these nanomaterials limit their use in biomedical and food packaging applications [16]. PLA nanocomposites reinforced with metallic and metallic oxide nanoparticles, such as silver, gold, copper, silica, and titanium oxide, not only enhance mechanical properties but also confer additional benefits, including improved antimicrobial and antiviral activity, thermal stability, and an increased glass transition temperature [17] . Among these, silver nanoparticles are of great interest as they enhance the mechanical properties of PLA, such as Young's modulus [18], tensile strength, toughness [19-22] and fracture properties [23], while also exhibiting strong antimicrobial properties [24]. These attributes make silver nanoparticle-reinforced PLA nanocomposites highly suitable for diverse applications in biomedical fields, including tissue engineering [25-27], food packaging [28], and cosmetics and hygiene products [29]. Furthermore, silver nanoparticles offer additional advantages such as enhanced thermal stability and low volatility [30].

In addition to surface coating, the architectural arrangement of nanofibers within a network significantly impacts its ultimate functionality. The specific arrangement of fibers profoundly influences their mechanical properties [31-33], adhesion, proliferation, and penetration of the cells, and also release of biomolecules from the network [34]. Furthermore, the wound healing process is significantly affected by the architecture and structure of scaffolds [8]. Considerable effort has been devoted to control the alignment and patterning of PLA network by employing various collection devices or by manipulating the electrical field during the electrospinning procedure [34]. As a result, fibre deposition and structure architecture can be tuned, in the form of random, up to aligned orientation.

Therefore, many intricacies are present in the design of nanofibrous networks including chemical composition, surface coating and fibre arrangement which challenges the prediction of mechanical behaviour and strength. The ability to achieve tailored design and efficient utilisation of nanofibrous network hinges upon thorough modelling and simulation of their mechanical performance.

The primary sources of complexity in the design and functional prediction of nanofibrous network stem from distinct physics and scales: atomistic/nanoscales and microscale. The deformation of freestanding nanofibres is predominantly governed by atomic-scale phenomena, especially in the presence of nanoscopic particles, whereas fibre arrangements and interconnections play their roles at the microscale. Therefore, a reliable modelling and design protocol must consider the interconnected physics simultaneously, especially when dealing with intricate failure mechanisms like crack growth and fracture. In the current work, a multiscale hierarchical framework is adopted to address the gap between governing physics in the fracture of nanofibrous networks by seamlessly integrating atomistic simulations at the nanoscale and peridynamics at the microscale. At the nanoscale, Molecular Dynamics (MD) simulation serves as a powerful tool to unravel the atomistic intricacies that dictate the fracture of freestanding pristine and silver doped polylactic acid nanofibres. In MD simulations, the interactions between the atoms and molecules are calculated using Newton's laws of motion, allowing for close analyses of atomistic phenomena and deformation mechanism, such as slip, twinning, and fracture that govern the bulk mechanical behaviour [35, 36]. The work presented in [37] is among the early reports in addressing fracture mechanics using MD simulations. The authors have applied axial load on a single crystal of $Al_2O_3$ with a centre pre-crack. To find the fracture toughness, the plot of energy versus crack length is fitted with Griffith criteria to find the critical crack length and the relevant stress. Some other works have focused on propagating crack speeds in single crystal silicone [38-40]. In a work by Deng et al. [41], MD was used to find the fracture toughness of a typical single crystal metallic glass. The fracture energy was considered as the mechanical work on the pristine sample up to the critical length from which the crack started growing in a notched sample. Deng et al. [42] further studied how the fracture toughness and crack growth in glass-ceramics (aluminosilicate) differs with the presence of crystalline circular or ellipsoidal regions inside the material [42]. The effect of hexagonal crystallised inclusions on fracture toughness of aluminosilicate glass-ceramic was also studied in [43]. The use of MD for addressing fracture mechanics is more common when it comes to graphene layers [44-47] or graphene-like layers [48]. Another approach to find the fracture toughness using MD was proposed in [49] and further applied by other researchers, such as [50] which used the external work on a centre pre-crack cell between the start of loading and total separation. In this work, it was assumed that at these two points, the strain energy was zero and the released energy was equal to the external work during this interval.

However, in our recent works on the tensile load on freestanding nanofibres, we proposed another approach to find the energy release rate [51, 52]. It was shown that after total separation of polymer nanofibres, the strain energy cannot be assumed as zero considering the stored energy due the rearrangement and plastic deformation of nanofibres. We have observed that prior to crack initiation, all the work done on the nanofibres is stored as strain energy, however, as the polymer chains begin to pull out, a growing difference arises between the external work and the stored strain energy, associated with the energy release due to crack propagation. This method is also used in the current work to find the fracture properties at the atomistic scale.

Despite the insights provided by atomistic simulation, its implementation at even meso scales are impractical due to the high computational cost. Therefore, as reviewed above, the studies on fracture properties using MD are mainly limited to single crystal cells or simple structures, such as graphene.

To overcome these computational limitations, multiscale strategies are used to bridge the simulation techniques across a range of length and time scales. Through a multi-scale nano-micro modelling, we can preserve the accuracy of atomistic modelling while benefiting from the efficiency of non-classical/non-local continuum description [53-55]. The challenge for a hierarchical multiscale method is to migrate seamlessly across scales so that the calculated properties can be efficiently transferred.

The choice of the microscale as the upper scale in the present hierarchical modelling approach is consistent with the multiscale nature of nanofibrous networks. Nanofibrous networks are composed of individual nanofibres, which are typically on the nanoscale, and the network structure usually emerges from the arrangement and interactions of these nanofibres at the microscale. Therefore, modelling the fracture behaviour at the microscale provides a natural transition from the atomistic scale while capturing the collective behaviour of the nanofibres within the network. On the other hand, choosing a higher scale such as macroscale within a one-level hierarchical approach may not be reasonable as increasing the scale gap between the two levels of modelling may lead to loss of information and overlooking critical factors at the microscale, such as configuration, connectivity, defects and surface effects. The adopted continuum mechanics at the upper scale, defined as microscale, should be capable of describing fracture and crack propagation in a material; however, the discontinuous nature of a crack contradicts basic hypothesis of Classical Continuum (CC) theory. To deal with this limitation and address crack formation and propagation, some methods have been developed to equip numerical methods that are based on CC, such as extended finite element method [56], partition of unity finite element method [57], element erosion [58], phase field model [59], and interface elements with a cohesive zone model [60]. Among continuum descriptors, Peri-Dynamic (PD) theory has shown proficiency in describing complex problems involving discontinuities, such as damage initiation, crack propagation and material interfaces. PD is in fact a non-local reformulation of CC mechanics based on integro-differential equations [61]. Given that the theory deals with integral equations rather than spatial differentiation in classical continuum, the PD governing equations are still valid in the presence of discontinuous displacement fields, such as crack surfaces allowing fracture and failure to be dealt with as natural material responses [62]. PD introduces a concept of damage for a material point, allowing to predict the evolution of cracks, including their nucleation, their propagation direction, and the points where they initiate and terminate, without excess criteria for triggering, bifurcation, and deviation [63]. PD is utilised to model several material systems, including polycrystalline materials [64], polymer composites [65-69], ductile metals [70], ceramics [71] and concrete [72]. PD has been employed for the analysis of impact [73], fatigue [74], and dynamic fracture [75-77], including determining stress intensity factor and fracture toughness of materials [78-83].

Therefore, considering the competence of PD to address fracture and crack propagation, it is implemented in the current work at the microscale to study the fracture of random and aligned fibrous networks. In a work by Bobaru [84], the tensile strength of a random nanofibrous network was determined from the stress-strain curve obtained from PD. Bobaru used the mechanical properties of the bulk polymer for the nanofibres while according to experimental observations and atomistic simulations the properties of the nanofibres differ from those of the bulk polymer and factors such as diameter and the surface coating will affect the final mechanical behaviour [51, 52]. Therefore, in the current work, an attempt is made to equip PD with material parameters of nanofibres that are determined from MD simulations.

Multiscale methods integrating MD and PD are efficient strategies, leveraging the precision of MD in capturing the atomic-scale fracture mechanism and efficiency of PD in describing fracture and crack propagation. In the current literature, MD and PD are incorporated within concurrent [85, 86] and hierarchical multiscale frameworks to study the mechanical response in several materials [87-90]. Gur et al. [87] applied the hierarchical method to upscale MD results for single crystals and bi-crystals of various orientations containing different grain boundaries by feeding it to the PD model to study elastic and fracture properties of polycrystalline 3C-SiC. At the atomistic scale, the energy release rate was obtained from the tensile load on a centre-cracked plate. In another work [89], the same methodology was employed to ascertain the stress-strain curve of Geopolymer Composites (GC) comprising calcium silicate hydrate, quartz, and the binder. Furthermore, multiscale MD-PD was employed in exploring the fracture of concrete gel with capillary pores [88], and single polyethylene microfibrils [90]. However, to the best of our knowledge, in the current literature no studies have established the linkage of MD and PD to investigate the mechanical and fracture properties of nanofibrous networks.

In the current work, we propose a hierarchical approach integrating MD at the atomistic scale and PD at the microscale to study fracture phenomena in aligned and randomly oriented nanofibrous networks. At the nanoscale, MD simulations are implemented to apply the tensile load on freestanding pristine and silver doped PLA nanofibres where the Young's modulus, Poisson's ratio and critical energy release rate are determined through innovative approaches. The obtained properties from MD will be transferred to inform PD at the micro scale to study the Young's modulus, fracture toughness of mode I and II and crack propagation in PLA nanofibrous networks.

A seamless data transition from MD to PD is ensured by the convergence of the tensile response of a single fibre in both descriptions. The proposed framework can bridge the gap between the governing physics in the fracture of nanofibrous networks at the nano to microscales. The method can account for the effect of surface coating and fibre arrangements on the measured properties.

It should be noted that in terms of implementation algorithms, coupled models can be solved in either a monolithic [91, 92] or staggered scheme [93]. In the staggered or partitioned scheme, the coupled equations are solved separately in a sequential manner. First, one set of equations (e.g., the MD equations) are solved, and then the results are used as input for solving the other set of equations (e.g., the PD equations). The linkage between the two sets of equations can be either non-iterative (static) or iterative (dynamic). In a monolithic or fully coupled scheme, the coupled equations are solved simultaneously as a single system. For instance, the MD and PD equations are integrated into a single set of equations that are solved together at each time step. This approach typically requires solving a larger, combined system of equations; however, it may ensure better consistency between the coupled fields. In the current study, the approach we have adopted can be classified as the staggered scheme. We begin by performing MD simulations to accurately capture the atomic-scale interactions and obtain elastic and fracture parameters. The PD simulations then use these static inputs to solve the continuum-scale peridynamic equations. This staggered approach allows us to take advantage of the detailed atomic-scale information provided by MD and apply it to the efficient continuum-scale PD, ensuring that the material behaviour is accurately represented across different scales. The separation of the two simulation steps simplifies the computational process and allows for the efficient handling of large-scale systems.

The rest of the paper is organised as follows: Section 2 describes MD simulation on freestanding pristine and silver doped PLA nanofibres and how to extract the required parameters to inform the higher scale. In Section 3, first the fundamentals of PD are reviewed, and then details are given on the hierarchical transition used to move from MD to PD. Further, the construction of aligned and randomly distributed fibrous network and the PD implementation to find fracture properties are addressed. The results on the elastic and fracture properties as well as crack propagation are presented in Section 4 with addressing the effect of fibre orientation and surface treatment on the obtained parameters. The concluding remarks are summarised in Section 5.

## 2. Molecular Dynamics Simulation of Untreated and Doped Nanofibres

### 2.1. Geometry

To find elastic and fracture properties of individual nanofibres required to inform peridynamics, molecular dynamics simulations have been conducted using LAMMPS [94]. The Open Visualization Tool (OVITO) developed by Stukowski et al. [95] has been used for rendering, and the initial structures were created in Materials Studio software [96]. Nanofibres are made of PLA with the molecular formula of $(C_3H_4O_2)_n$ [97] as shown in Fig. 1a. Silver nanoparticles (Ag-NP) with a diameter of 1.8 nm are used as doping agent of nanofibres shown in Fig. 1b. Regarding the choice of the Ag-NP diameter, we should note that different studies have reported on the size dependent antibacterial features of silver nanoparticles [98-102] with smaller particles exhibiting higher antibacterial activity. Specifically, in a renowned work by Morones et al [24], the authors have studied the antibacterial properties of silver nanoparticles ranging from 1 to 100 nm and they have reported the best enhancement is attributed to silver nanoparticles with diameter of ∼1–10 nm. Therefore, in the current work, we chose to simulate a nanoparticle with a diameter of 1.8 nm. Opting for a small nanoparticle is also in favour of less computational cost.

According to Fig. 1c, the polymer simulated in the current work is PDLA stereoisomer of PLA due to the relative position of the methyl group to the chain backbone [103, 104]. In contrast to PLLA stereoisomer, PDLA has no tendency for crystallization [105]. Therefore, in the current work, nanofibres are in the amorphous state.

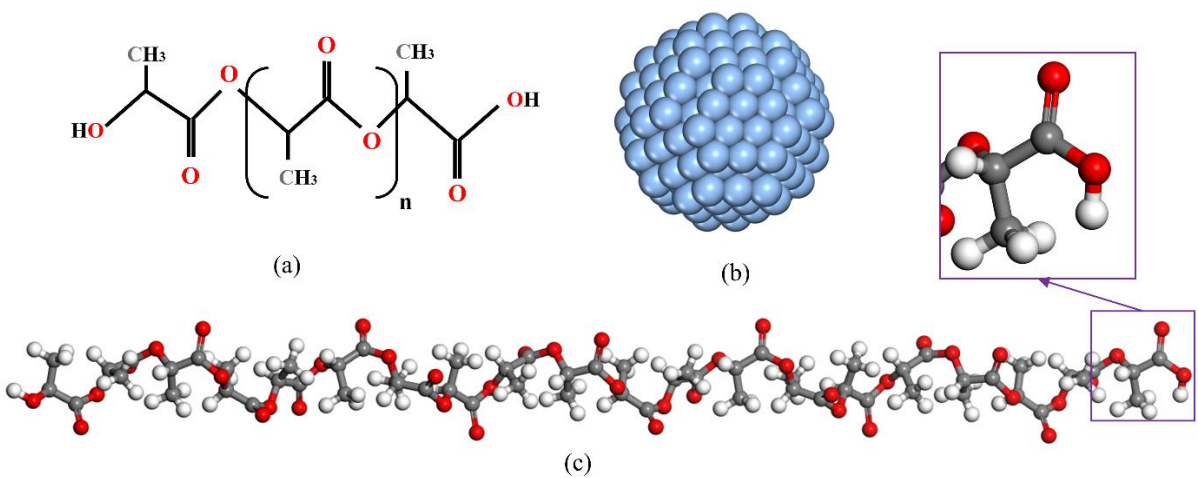

Fig. 1 (a) Chemical structure of PLA, (b) Silver nanoparticle, and (c) PDLA stereoisomer of PLA

To construct polymer nanofibres (PNFs), the method described in detail in [51, 52] was followed. Briefly addressing, after equilibrium of a cuboidal simulation box consisting of 90 numbers of 20-monomer PLA chains (Fig. 2a), a relaxation procedure, detailed in [51] and [106], is followed where a canonical NVT ensemble at 500 K (well above the glass transition temperature) and a further isothermal-isobaric (NPT) ensemble at 500 K and 1 atm are conducted on the simulation box for 5 nanoseconds (ns). The system is then cooled down to 298 K within 10 ns through NPT ensemble. In an NVT ensemble, the total number of atoms, volume of the simulation box, and the temperature of the system are kept constant while in an NPT ensemble, the atoms number and the system pressure and temperature are maintained.

Further, the *x*- and *y*-dimensions of the box are expanded by 5-7 times the cut-off radius without rescaling the atomic coordinates, while the dimension in the z-direction remains unchanged (Fig. 2b). In this way, a nanofibre with a diameter of 5.4 nm is obtained, which is isolated in the transverse direction and has an infinite axial length. The silver nanoparticles are then manually positioned on the surface of the nanofibres, followed by energy optimisation and relaxation through a canonical ensemble (Fig. 2c).

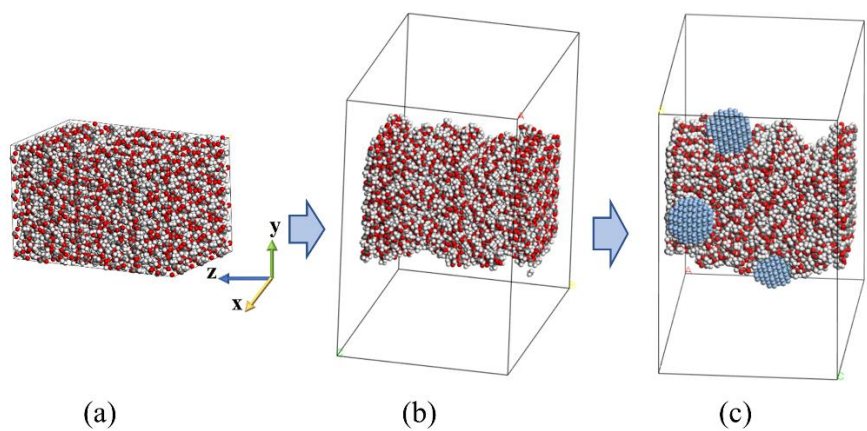

(a)          (b)          (c)

Fig. 2-a) Initial cuboid, b) relaxed nanofibre in the enlarged cell, and c) doping silver nanoparticles on nanofibre

## 2.2. Interatomic Potential

In this study, all-atom simulations are employed instead of the coarse-grained approach to enhance the precision and reliability of the results. The coarse-grained method simplifies computations by treating groups of atoms as single beads and ignoring internal interactions, which, while reducing computational cost, decreases precision. The ab initio Class II polymer consistent forcefield (PCFF) is assigned to represent atomic interactions within the polymer since it is adept at modelling polymers and organic materials [107, 108]. PCFF is well-known for its ability to characterize cohesive energies, mechanical characteristics, compressibility, heat capacities, and elastic constants [109]. The total energy in PCFF ($E_{PCFF}$) is divided into valence\bonded ($E_{valence}$) and non-bonded interactions ($E_{non-bonded}$) according to Equation (1-4):

$$E_{PCFF} = E_{valence} + E_{non-bonded}, \tag{1}$$

Where valence interactions include bond ($E_{bond}$), angle ($E_{angle}$), dihedral ($E_{dihedral}$), inversion ($E_{inversion}$), and the cross-coupling ($E_{cross-coupling}$) interactions between atoms:

$$E_{valence} = E_{bond} + E_{angle} + E_{dihedral} + E_{inversion} + E_{cross-coupling}. \tag{2}$$

with bond and angle interactions incorporating quartic terms to capture anharmonic properties. Torsion interactions are modelled using a symmetric Fourier expansion, while out-of-plane contributions are represented by a simple harmonic function [110].

$$E_{bond} = \sum_{n=2}^{4} k_n^b (b - b_0)^n, \; E_{angle} = \sum_{n=2}^{4} k_n^\theta (\theta - \theta_0)^n,$$

$$E_{dihedral} = \sum_{n=2}^{4} k_n^\phi (1 - \cos(n\phi))^n, \; E_{inversion} = k^\chi (\chi - \chi_0)^2, \tag{3}$$

where $k_i^j$ represents i-th coefficient related to j-th kind of energy where the superscripts b, θ, ϕ and χ are assigned for bond, angle, dihedral and inversion. More information on the cross-coupling term can be found in [110].

The non-bonded term includes electrostatic ($E_{electrostatic}$) interactions and a 9–6 Lennard–Jones potential to account for the van der Waals forces ($E_{van\,der\,Waals}$).

$$E_{non-bonded} = E_{electrostatic} + E_{van\,der\,Waals},$$

$$E_{van\,der\,Waals} = \epsilon \left[ \left(\frac{\sigma}{r}\right)^9 - \left(\frac{\sigma}{r}\right)^6 \right] \quad r < r_{cutoff}, \tag{4}$$

$$E_{electrostatic} = q_i q_j / r,$$

where r is the distance between an atom pair, ϵ and σ are the well depth and zero-crossing distance and $q_i$ is the charge of atom i. For all simulations, the cut-off radius is $r_{cutoff}$ = 1.05 nm [111].

The embedded-atom method (EAM) potentials [112] is used to model the interactions between silver atoms within the nanoparticle. EAM has been tested by calculating a wide range of properties such as migration energy of vacancies, the formation energy, migration energy of divacancies and self-interstitials, the surface energy and geometries of the low-index surface of the pure metals. For a system consisting of N atoms, the EAM potential will be:

$$E_{EAM} = \frac{1}{2}\sum_{i=1}^{N-1}\sum_{j=i+1}^{N}\varphi(r_{ij}) + \sum_{i=1}^{N}F(\rho_i) \tag{5}$$

Where

$$\rho_i = \sum_{j=1, j\neq i}^{N}\psi(r_{ij}) \tag{6}$$

$\varphi(r_{ij})$ is the pairwise repulsive energy between atoms i and j and F is the energy required to embed an atom I in a point with local electron density $\rho_i$. The local electron density is described by superposition of contributions $\psi(r_{ij})$ from neighboring atoms.

For the cross-terms between polymer and nanoparticle atoms, we have used the 6th order combination law proposed by [113] and recommended by the well-known work of Sun [114] for unlike atom pairs:

$$\epsilon_{ij} = 2\sqrt{\epsilon_i\epsilon_j}\frac{\sigma_i^3\sigma_j^3}{\sigma_i^6 + \sigma_j^6} \quad r < r_{cutoff},$$

$$\sigma_{ij} = \left(\frac{1}{2}(\sigma_i^6 + \sigma_j^6)\right)^{\frac{1}{6}} \quad r < r_{cutoff} \tag{7}$$

For 9–6 Lennard–Jones potential of silver atoms $\epsilon = 4.1$ and $\sigma = 3.0222$. These numbers are available on the PCFF force field parameter files in LAMMPS which are retrieved from [115] and [116].

2.3. Derivation of Material Parameters

As detailed in Section 3.1, to inform prototype micro elastic brittle peridynamics, the two required material parameters are the bulk modulus ($K$) and the critical energy release rate ($G_0$). To find these material parameters, a uniaxial deformation with a constant strain rate of $10^8$ 1/s was applied on each nanofibre by uniformly changing the box size in the fibre's axial direction (Fig. 4a). Considering that in MD simulations the time and length scales are in the femtosecond and nano meter ranges, a strain rate ranging from $10^5$ 1/s - $10^{10}$ 1/s is typically applied to deformations [117]. The effect of strain rate on the material parameters of PLA nanofibres is addressed in [51].

The Langevin thermostat [118] within the microcanonical ensemble (NVE) is used during deformation. As the lateral surface is unconstrained, the deformation resembles a uniaxial tensile test. The velocity-Verlet algorithm was applied to integrate Newton's equations of motion with a time step of 0.5 femtoseconds. It should be noted that in LAMMPS, in case of using canonical (NVT), isothermal-isobaric (NPT), and isenthalpic (NPH) ensembles, the time integration is performed with Nose-Hoover style non-Hamiltonian equations of motion which are designed to generate positions and velocities. Therefore, it performs time integration as well as Nose/Hoover thermosstatting. However, in case of using Langevin thermostat, the thermostat only modifies forces to account for the effect thermostatting. Thus, we must use a separate time integration, like NVE ensemble, to update the velocities and positions of atoms using the modified forces. The use of Langevin thermostat, as described in [118] was to set the temperature of the atoms while modelling the interaction with an implicit background. The Langevin thermostat and the NVE ensemble is performed in conjugation with strain application.

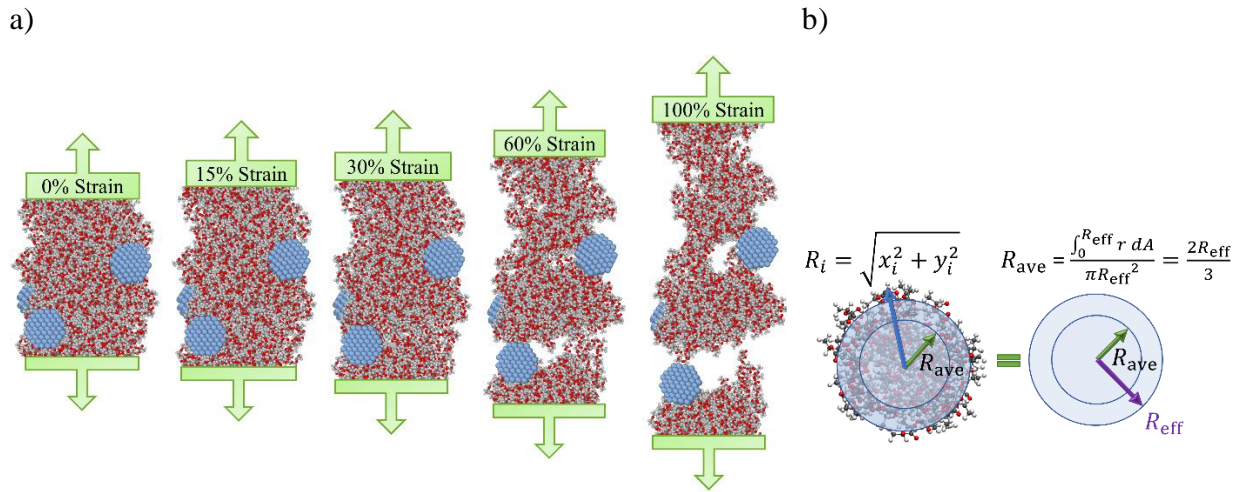

Fig. 3 a) Tensile deformation on the nanofibre doped with 5.8% weight fraction of silver nanoparticles, and b) determination of effective radius where the average radial distance ($R_i$) of all atoms from the centroid is set equal to the mean radius ($R_{ave}$) of an equivalent solid circle

The stress tensors are calculated using the Virial theorem [119]. To calculate the Virial stress, the definition of the nanofibre cross-section is required. Therefore, we propose a method illustrated in Fig.3b for the equivalent radius which results in less fluctuation during uniaxial deformation compared to the Gibbs dividing surface [120-123]. At each time, the average radial distance ($R_i$) of all atoms from the centroid is set equal to the mean radius ($R_{ave}$) of an equivalent solid circle. The radius of this equivalent cylinder is supposed as the effective radius ($R_{eff}$) of nanofibres considering $R_{eff} = 3R_{ave}/2$.

We should note that to calculate the axial force, the stress tensors are first calculated using the Virial theorem [119]. However, in LAMMPS the applied volume in the denominator of the Virial stress is the volume of the simulation box. Thus, to find the axial force, the Virial stress is multiplied by the length of the simulation box in the fiber direction.

The bulk modulus ($K$) is calculated using the Young's modulus ($E$) and Poisson's ratio ($\nu$) of nanofibres through the well-known relation $K = E/3(1 - 2\nu)$. The Young's modulus is determined using the slope of the linear elastic regime in the stress-strain curve (Fig. 4), and Poisson's ratio ($\nu$) is determined using the radial deformation in the elastic regime considering $R_{eff} = R_0(1 - \nu\epsilon_z)$ where $R_0$ is the initial effective radius and $\epsilon_z$ is the axial strain (Fig. 5).

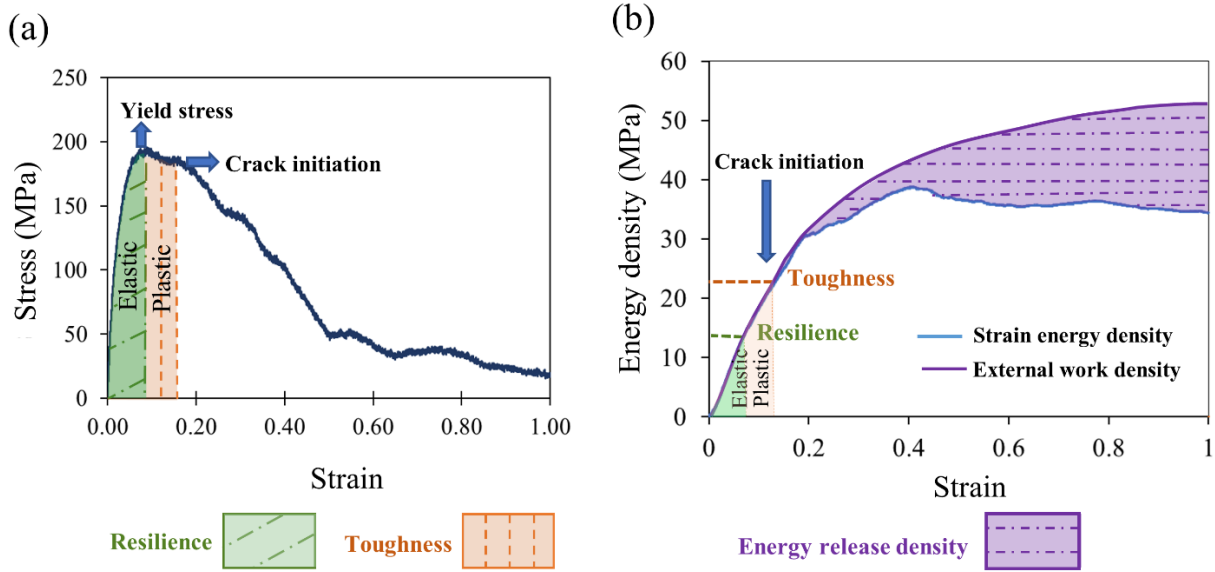

Fig. 4 (a) Stress-strain response of silver doped PLA nanofibre diameter deformed in uniaxial tension, and (b) Evolution of strain energy and external work densities

The critical energy release rate ($G_0$) which measures the amount of energy required to grow a crack is calculated from the difference between the elastic energy stored in the material and the energy required to create new surfaces as the crack extends[124].

As shown in Fig. 4b, prior to crack initiation, all the work done on the nanofibre is stored as the strain energy; however, as the polymer chains begin to pull out, a growing difference arises between the external work ($W_{ext}$) and the stored strain energy ($U$), associated with the energy release due to crack propagation. The method to find the energy release rate in the current work is illustrated in Fig. 5. First, the cross-sectional area is calculated just before crack initiation ($A_{cr}$). The crack area is thus equal to zero at crack initiation and it equals to $A_{cr}$ at total separation. The energy release rate equals the difference $U - W_{ext}$ at these two points divided by $A_{cr}$. $W_{ext}$ is calculated as $\int_0^l F dl$ by integrating of an interpolation function for force ($F$)-displacement ($l$) data. The strain energy is obtained from the instant potential energy subtracted from its initial value.

It should be clarified that in the estimation of the energy release rate, the change in the cross-sectional area of the nanofibre is taken into account; The cross-section of the nanofibre is dynamically calculated using the effective radius as shown in Fig. 5. As illustrated in Fig. 6, $A_{crack}$ is equal to zero where the crack initiates and equal to the nanofiber area where the full detachment occurs. The other two parameters involved in the estimation of the energy release rate, $W_{ext}$ and $U$ are defined independent of the cross-section; $W_{ext}$ is determined by integrating the force-displacement curve, where the axial force is derived from the Virial stress, multiplied by the length of the simulation box in the fiber direction. The potential energy, $U$ is calculated directly as the difference between the instantaneous potential energy and its initial value.

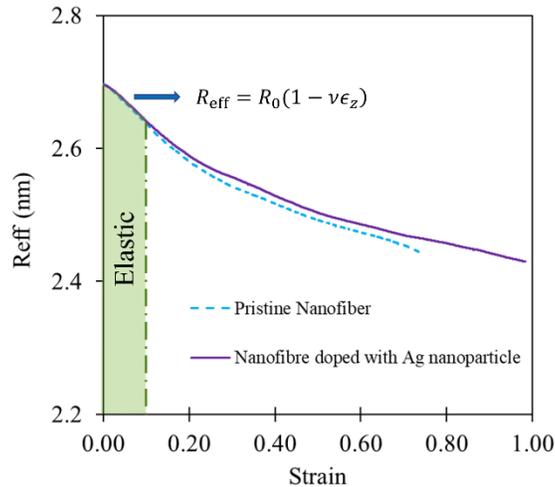

Fig. 5 The change in effective radius during deformation and determination of Poisson's ratio. The determined mechanical parameters for pristine and doped nanofibres are listed in Table 1.

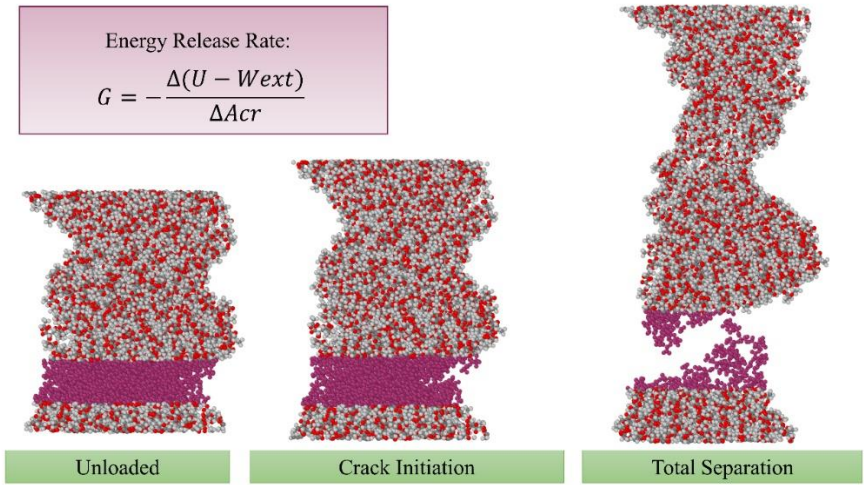

Fig. 6. Determination of energy release rate for nanofibres. The energy release rate equals the difference of $U - W_{ext}$ divided by $A_{cr}$ between the total separation and the onset of crack propagation

Table 1. Mechanical parameters of pristine and silver-doped nanofibres from MD simulations

|  | E (GPa) | ν | K (GPa) | $G_0$ (m J/m²) |
|---|---|---|---|---|
| Pristine Nanofibre | 2.6 | 0.22 | 1.542 | 365 |
| Nanofibre doped with Ag nanoparticle of 3.9 % weight fraction | 2.8 | 0.19 | 1.503 | 496 |

3. Peridynamics Theory and Implementation

   3.1. Fundamentals of Peridynamics

In the current work, two different descriptors are used at nanoscale and microscale. At the nanoscale, molecular dynamics simulation is implemented to describe atomic scale phenomena, while at the micro scale, peridynamics simulations are applied. For the sake of brevity and since MD is well established [125-127], this subsection gives an overview on the fundamentals of PD.

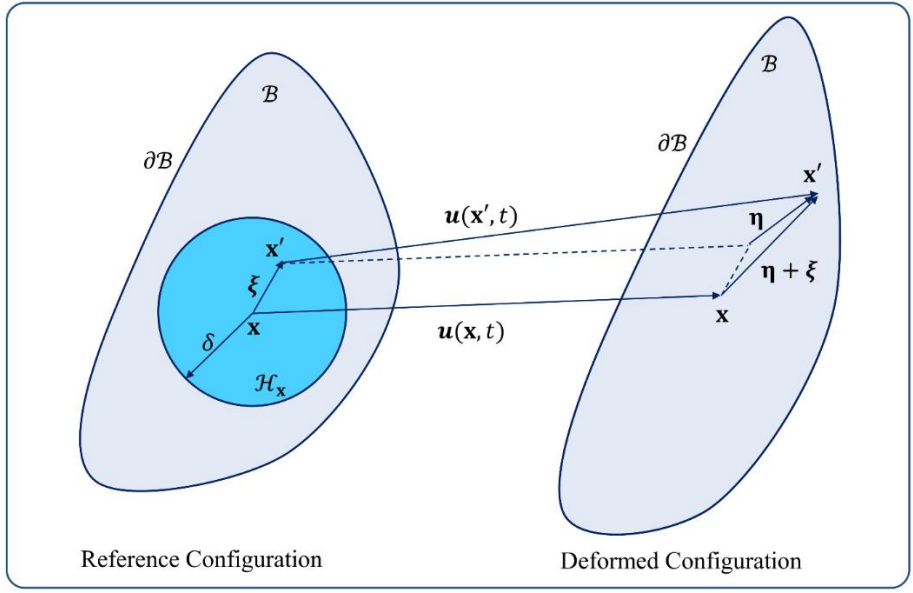

Fig. 7 Schematic representation of a generic PD domain $\mathcal{B}$ before and after deformation; the relative position vectors and the relative displacement vector between the material points $x$ and $x'$ are shown [62]

The peridynamic theory is a non-local reformulation of classical continuum mechanics based on integro-differential equations. Considering that PD deals with integral equations rather than spatial differentiation, PD governing equations are valid even in presence of discontinuous displacement fields, such as cracks, and therefore it is suitable for damage prediction in dynamic fracture analysis. As shown in Fig. 7, in the PD framework, in a domain $\mathcal{B} \subset \mathbb{R}^3$, described with PD, each material point, **x**, belonging to a domain $\mathcal{B} \subset \mathbb{R}^3$ interacts with all the other material points located within a finite neighbourhood referred as $\mathcal{H}_x$. For any material point $\mathbf{x} \in \mathcal{B}$ at time $t \geq 0$, the bond-based PD equation of motion is given by [128]:

$$\rho(\mathbf{x})\ddot{\mathbf{u}}(\mathbf{x},t) = \int_{\mathcal{H}_x} \mathbf{f}(\mathbf{u}(\mathbf{x}',t) - \mathbf{u}(\mathbf{x},t), \mathbf{x}' - \mathbf{x}) dV_{\mathbf{x}'} + \mathbf{b}(\mathbf{x},t), \qquad (8)$$

where $\rho(\mathbf{x})$ is the mass density and $\ddot{\mathbf{u}}$ is the second time derivative of the displacement field $\mathbf{u}$. $\mathbf{f}$ is the pairwise force function, with units of force per unit volume squared, which represents the micro-force density that the material point $\mathbf{x}'$ exerts on the material point $\mathbf{x}$. $dV_{\mathbf{x}'}$ is the infinitesimal volume associated with the material point $\mathbf{x}'$, and $\mathbf{b}$ is a prescribed body force density field. The neighbourhood $\mathcal{H}_\mathbf{x}$ is defined through the following relation:

$$\mathcal{H}_x := \{\mathbf{x}' \in \mathcal{B} : \|\mathbf{x}' - \mathbf{x}\| \leq \delta\}, \qquad (9)$$

where $\delta > 0$ is called the horizon. For a material point in the bulk of a 3-dimensional body, the neighbourhood $\mathcal{H}_\mathbf{x}$ is a spherical integration region centred at $\mathbf{x}$. As shown in Fig. 7, the relative position vector between the material points $\mathbf{x}$ and $\mathbf{x}'$ in the reference configuration, which represents the standard notation for a bond, is defined as:

$$\boldsymbol{\xi} := \mathbf{x}' - \mathbf{x}. \tag{10}$$

The relative displacement vector of the two material points $\mathbf{x}$ and $\mathbf{x}'$ in the deformed configuration is instead denoted by:

$$\boldsymbol{\eta} := \boldsymbol{u}(\mathbf{x}', t) - \boldsymbol{u}(\mathbf{x}, t), \tag{11}$$

where $\boldsymbol{u}(\mathbf{x}', t)$ and $\boldsymbol{u}(\mathbf{x}, t)$ represent the displacements of the material points $\mathbf{x}'$ and $\mathbf{x}$ at time $t > 0$, respectively. The relative position vector of the two material points in the deformed configuration is then denoted as $(\boldsymbol{\eta} + \boldsymbol{\xi})$. The force vector $\mathbf{f}$, which contains all the constitutive information of the material, acts in the direction of $(\boldsymbol{\eta} + \boldsymbol{\xi})$. The general form of $\mathbf{f}$, which is commonly referred to as bond force, can be written as:

$$\mathbf{f}(\boldsymbol{\eta}, \boldsymbol{\xi}) = f(\boldsymbol{\eta}, \boldsymbol{\xi}) \frac{\boldsymbol{\eta} + \boldsymbol{\xi}}{\|\boldsymbol{\eta} + \boldsymbol{\xi}\|}, \tag{12}$$

where $f(\boldsymbol{\eta}, \boldsymbol{\xi})$ is a scalar even function for which the value depends on the material type. In the Prototype Microelastic Brittle (PMB) material model introduced in [61], the force vector $\mathbf{f}$ is derived from a scalar differentiable function $w$, referred to as pairwise potential function, such that:

$$f(\boldsymbol{\eta}, \boldsymbol{\xi}) = \frac{\partial w(\boldsymbol{\eta}, \boldsymbol{\xi})}{\partial \boldsymbol{\eta}}, \tag{13}$$

where $w$ takes the following form:

$$w(\boldsymbol{\eta}, \boldsymbol{\xi}) = \frac{cs^2 \|\boldsymbol{\xi}\|}{2}, \tag{14}$$

in which $c$ is referred to as micromodulus function and represents the bond elastic stiffness, and $s$ is the bond stretch, expressed by:

$$s = \frac{\|\boldsymbol{\eta} + \boldsymbol{\xi}\| - \|\boldsymbol{\xi}\|}{\|\boldsymbol{\xi}\|}. \tag{15}$$

For the generalized PMB material model, $f(\boldsymbol{\eta}, \boldsymbol{\xi})$ is therefore given by:

$$f(\pmb{\eta}, \pmb{\xi}) = cs. \tag{16}$$

As comprehensively discussed in [61, 72, 129, 130], through the following relation, the micromodulus $c$ can be related to the bulk modulus, $K$, which is a measurable macroscopic quantity:

$$c = \frac{18K}{\pi \delta^4} \tag{17}$$

In the PMB material model, a critical value for the bond stretch, $s_0$, is defined in order to introduce the concept of material failure [61]. As soon as the stretch $s$ of a bond exceeds this limit value, the bond is considered broken, and it can no longer withstand the tensile force. In this regard, Eq. (15), can be written as [128]:

$$\mathbf{f}(\pmb{\eta}, \pmb{\xi}) = \mu(\pmb{\xi}, t) cs \frac{\pmb{\eta}+\pmb{\xi}}{\|\pmb{\eta}+\pmb{\xi}\|}, \tag{18}$$

where $\mu(\pmb{\xi}, t)$ is a history-dependent scalar function which is introduced as a bond-breaking parameter, and takes either of the following values:

$$\mu(\pmb{\xi}, t) = \begin{cases} 1 & if\ s(t') < s_0,\ \ 0 < t' < t \\ 0 & otherwise \end{cases} \tag{19}$$

The history-dependence of the constitutive model is due to the rupture of a bond being an irreversible process. The critical stretch can also be derived as a function of measurable macroscopic quantities, such as the critical energy release rate, $G_0$ of the material [61].

$$s_0 = \sqrt{\frac{5G_0}{9K\delta}} \tag{20}$$

However, as pointed out in [61], for some materials, such as glass, considering a variable critical stretch leads to better agreement with experimental results. For a more realistic estimation, $s_0$ at a given point will depend on the conditions of all the connecting bonds through the following relation:

$$s_0(t) = s_{00} - \alpha\, s_{min}(t) \tag{21}$$

where $s_{min}$ is the current minimum stretch among all bonds connected to a given material point:

$$s_{min}(t) = min_\xi \left\{ \frac{\|\boldsymbol{\eta} + \boldsymbol{\xi}\| - \|\boldsymbol{\xi}\|}{\|\boldsymbol{\xi}\|} \right\} \tag{22}$$

In this case, $s_{00} = \sqrt{\frac{5G_0}{9K\delta}}$ and α is a constant typically set as ¼.

Damage at a material point **x** at each time $t$ is expressed by introducing a local damage index, $\varphi(\mathbf{x}, t)$, defined by the following relation:

$$\varphi(\mathbf{x}, t) = 1 - \frac{\int_{\mathcal{H}_x} \mu(\boldsymbol{\xi}, t) dV_{x'}}{\int_{\mathcal{H}_x} dV_{x'}} \tag{23}$$

where $0 \leq \varphi(\mathbf{x}, t) \leq 1$, the value of 0 represents the undamaged state of the material, and 1 indicates the complete disconnection of the material point x from all the material points located within its neighbourhood [61].

### 3.2. Hierarchical Data Transition from Molecular Dynamics to Peridynamics

The PMB material model requires three material parameters - micro-modulus, critical bond stretch, and parameter α, typically set as ¼ [61]. According to Eq. (10) and (13), $c$ and $s_0$ are related to the horizon $\delta$ and the measurable macroscopic quantities bulk modulus $K$, and energy release rate $G_0$. In the current work, the values of constitutive parameters, $G_0$ and $K$, were determined for polymer nanofibres using MD simulations in Section 2. However, care must be taken to choose the proper value for the horizon. The horizon is often selected to be three times the grid space, but as pointed by the original work of Siling and Askari [61], $\delta$ and $c$ can be justified to match the experimental data. Furthermore, as mentioned in [131], the PMB model of bodies without pre-cracks yield different strengths when different horizons are used to simulate crack nucleation. The horizon refers to a characteristic length scale that defines the distance over which material points interact with each other. It plays the main role in non-local nature of PD to model phenomena such as crack initiation and propagation. On the other hand, horizon acts as a regularization parameter, smoothing out the stress and displacement fields near material discontinuities.

In this study, the horizon is calibrated to ensure that the maximum withstanding force for a single fibre simulated in PD matches the value obtained from MD simulation. The numerical implementation of PD is carried out using the PDLAMMPS package [132] in LAMMPS software. A single fibre of 500 μm is simulated in PDLAMMPS and a grid-spacing of 5 μm is used for discretization with a study to verify the grid-independence. The overall unit of micron is chosen in the LAMMPS script. Velocities of 1 μm /μs are applied over a length of 50 μm at each end of the fibre. To avoid an initial shock at the boundaries, a linear velocity distribution from +1 μm /μs to -1 μm /μs along the fibre direction is considered in the initial stages. The time step is set as $10^{-6}$ μs.

It should be noted that in some works, such as [89] and [87], the PD spacing is of the order of nm for hierarchical connection of MD and PD. However, this assumption has led to a significant difference in the predicted strength between MD and PD while the lattice spacing is a numerical parameter and there is no need to be restricted to this small order. In contrast, Zhang et al. [88] artificially increased the energy release rate to achieve convergence between the predicted values of the two disciplines. However, in the current work, by adjusting the horizon as a numerical parameter, the convergence is achieved while retaining the original material parameters.

Fig. 8a presents the force-displacement curve and configurations obtained from MD and PD after the calibration of the horizon value.

As seen in Fig. 8a, a proper choice of the horizon allows the ultimate force and strength of the nanofibre in PD to match those of MD. However, it should be noted that the crack initiates much earlier in PD than in MD. This limitation corresponds to the inherent nature of the PMB model, where the fracture parameter is related to the energy release rate for a growing crack rather than crack nucleation [131]. Despite this drawback, PMB has been shown to be effective to reproduce fracture behaviour with careful implementation [64-77]. In the current work, this limitation is addressed by scaling the strain of PD to match that of MD. The final input parameters for PD implementation are presented in Table 2.

Table 2. Obtained Parameters for PMB model implementation including $\delta$ (horizon), $c$ (micromodulus) and $s_0$ (critical stretch)

|  | horizon $\delta$ ($\mu m$) | micromodulus $c$ ($\frac{\text{pg}}{(\mu s)^2 \ (\mu m)^5}$) | critical stretch $s_0$ |
|---|---|---|---|
| Pristine Nanofibre | 60 | 0.6819 | 0.00148 |
| Nanofibre doped with Ag nanoparticle | 60 | 0.6643 | 0.00175 |

As inferred from Table 2, doped nanofibres have a lower micro-modulus and higher critical bond stretch, indicating increased ductility compared to untreated nanofibres. These differences are reflected in the obtained PD input parameters.

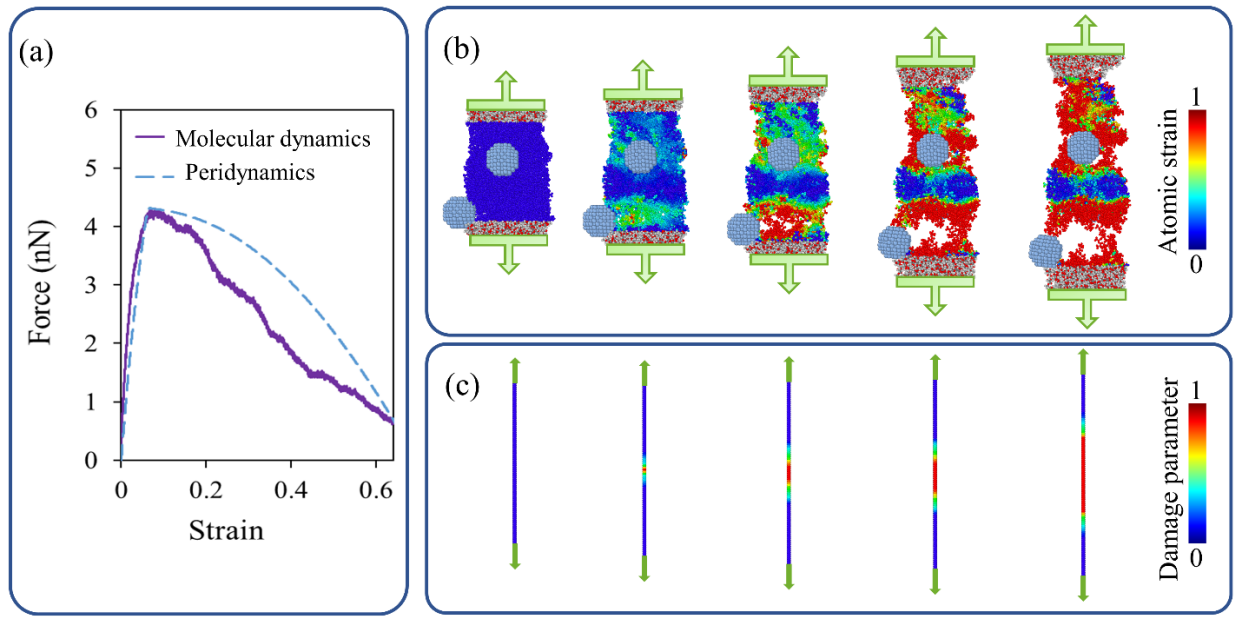

Fig. 8 a) The applied force comparison on an individual nanofibre using MD and PD implementations, b) The nanofibre's configurations simulated in MD where the colours indicate the atomic strain, and c) The nanofibre's configurations simulated in PD where the colours indicate the damage parameter, $\varphi$.

It should be clarified that the anisotropy within the nanofibre may have two origins, the intrinsic anisotropy due to variations in material properties in different directions, and the geometrical anisotropy due to the configuration of the nanofibre. In the current work, we are simulating a nanofibre made from a polymer with isotropic material properties and the geometrical anisotropy is taken into account as we are calibrating the fibre in MD with a single fibre in PD with the same configurational shape. As observed in the atomistic simulation, the polymer density and consequently the material properties vary along the radial direction of the nanofiber due to the presence of a surface layer. However, this variation occurs over a nanometre distance and does is not transferred to the upper scale in the coupling method.

### 3.3. PD Simulation of Random and Aligned Fibrous Network

### 3.4. Initial Construction of Fibrous Network

To construct the configuration for three-dimensional random fibrous network, a cuboid of 2H µm×2W µm×t µm dimensions is considered where a specified number of fibres, M, pass through each pair of planes of the cuboid, as shown in Fig. 9. The fibres connecting each pair of the planes are the lines connecting M random points on each plane to the other one. For each endpoint, one coordinate is already given from the plane on which it lies, and the other two coordinates are generated using the *rand* function in MATLAB, which creates uniformly distributed random numbers.

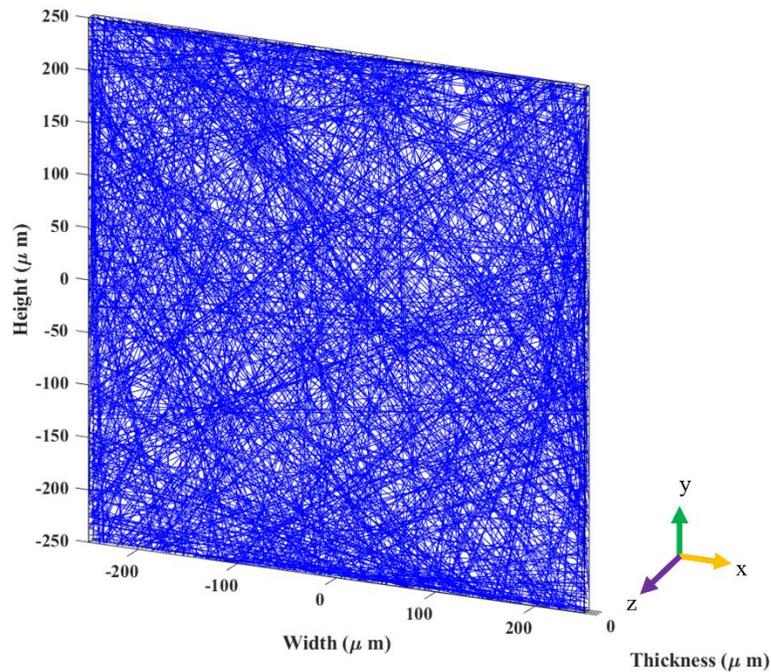

Fig. 9. 3D dimensional random fibrous network. (2L=500μm, 2W=500 μm, t=5 μm and M=150)

To prepare the pointwise configuration required for PD, the fibres are discretised into the evenly distributed points with the lattice spacing of 5 μm. Fig. 10a shows the discretised configuration of the network presented in Fig. 9. The colours vary depending on the $z$ coordinate. In Fig. 10b, SEM image of an electrospun PLA fibrous mat with randomly oriented fibres is presented for comparison.

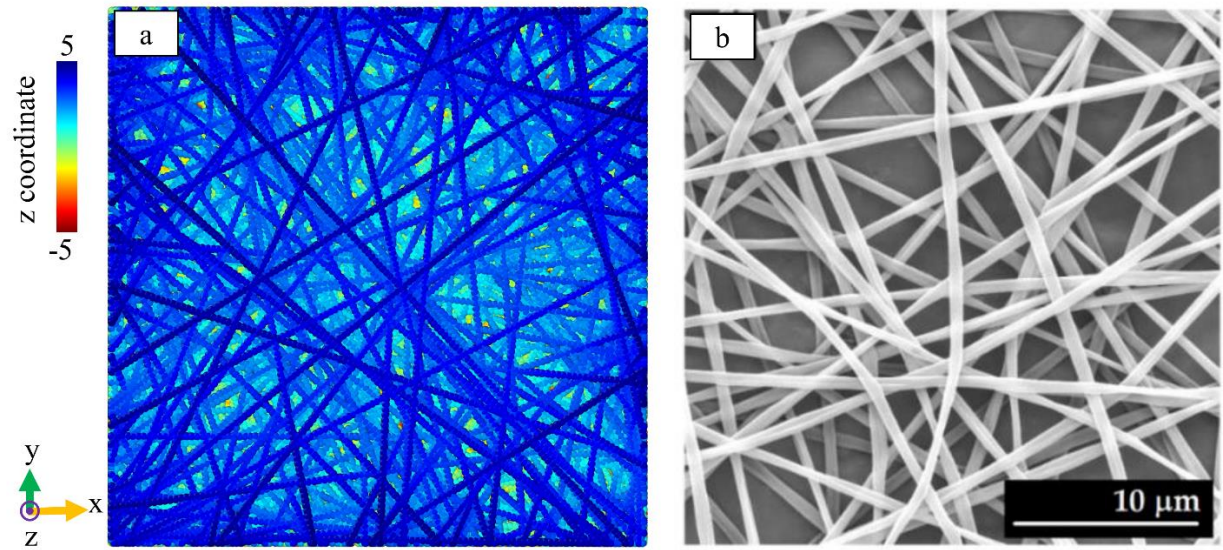

Fig. 10. a) Pointwise discretized configuration for the 3D dimensional random fibrous network, and b) SEM image of an electrospun PLA randomly oriented fibrous mat [34] (distributed under the terms of the Creative Commons)

The coordinates for the aligned fibrous network are generated using the MATLAB function *meshgrid* with the fibres aligned in the y direction, as shown in Fig. 11a. In Fig. 11b, for comparison, SEM image of an electrospun PLA fibrous mat with aligned fibres is presented.

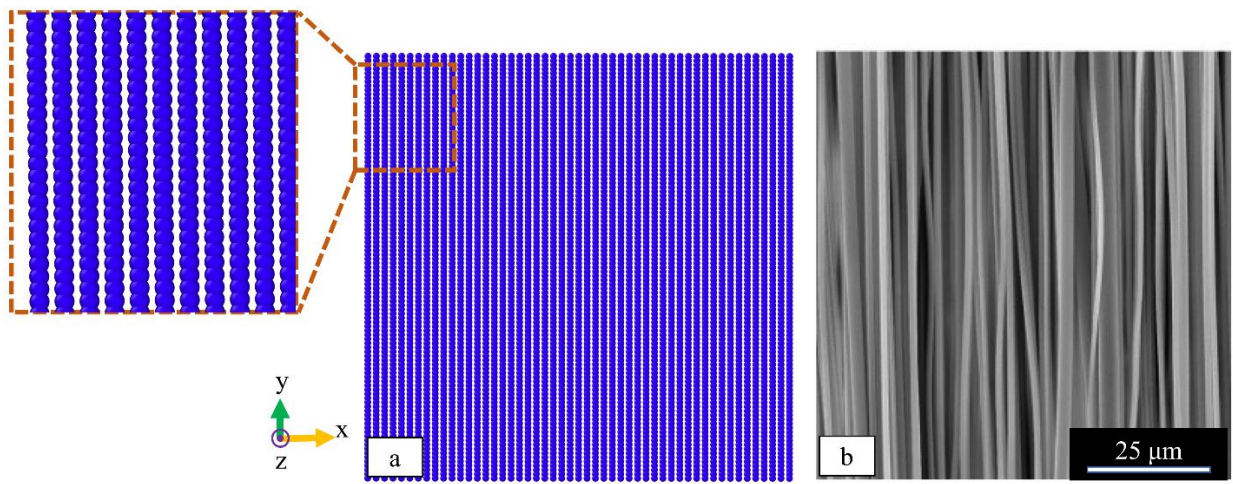

Fig. 11. A) Pointwise discretized configuration for the aligned fibrous network, and b) SEM image of an electrospun PLA aligned fibrous mat [133] (distributed under the terms of the Creative Commons CC)

### 3.5. PD Implementation

The same PD implementation settings for the individual nanofibre in Section 3.2 is adopted for the random and aligned fibrous networks. Velocities of 1 μm/μs are applied over a strip of 50 μm width on horizontal edges (Fig. 12) with an initial linear velocity distribution along the y direction. The time step is set as $10^{-6}$ μs. The forces applied on each of the strips are computed during deformation. A periodic simulation box is considered with the size of the box 4 times larger than the network dimensions to avoid interactions of the particles in image cells.

The stress is calculated from the obtained force divided by the network overall cross-section, and the Young's modulus is obtained from the slope of the linear part of the stress-strain curve. Further, an initial crack of length $2a = 200\ \mu m$ in the centre of the plate is considered. To account for the presence of the crack in the numerical implementation, the interaction between the particles located on either side of the crack is neglected. The maximum stress from the tensile test on the pre-cracked specimen is used to calculate the critical stress intensity factor as a measure of the fracture toughness [124]. The following empirical relation [134] for a centre-cracked finite plate is used to calculate $K_I$, the stress intensity factor related to mode I or opening mode of fracture, with about 1% error for any crack to plate width ratio.

$$K_I = \sigma\sqrt{\pi a}\left[\sec\left(\frac{\pi a}{2W}\right)^{1/2}\right]\left[1 - 0.025\left(\frac{a}{W}\right)^2 + 0.06\left(\frac{a}{W}\right)^4\right] \tag{24}$$

where 2a is the crack length and W is the plate width as indicated in Fig. 12.a.

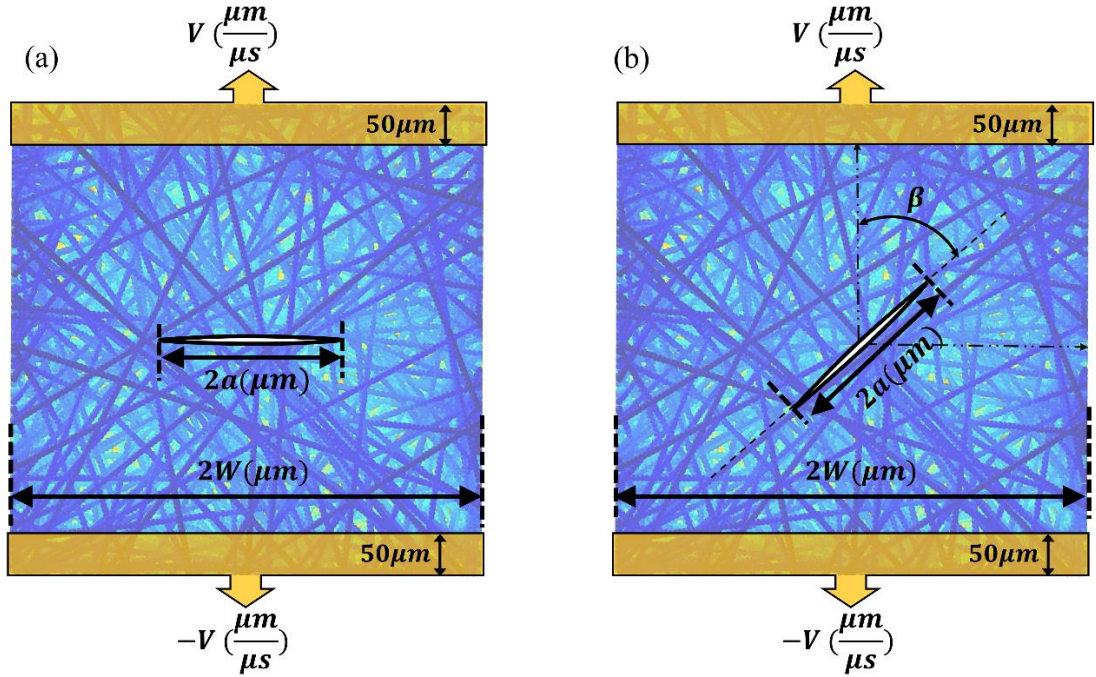

Fig. 12. Tensile load on a pre-cracked random fibrous network with the crack length of $2a$: a)

perpendicular to the load direction to represent mode I loading, and b) inclined crack at the angle $\beta$ from the load direction to represent mixed-mode I and II loading

To find the fracture toughness related to mode II or the shear mode, the tensile load is applied on the plate with an inclined crack at the angle $\beta = 60°$ from the load direction (as shown in Fig.12.b), which created a mixed-mode condition. Using the maximum stress in the case of inclined crack and having the opening-mode fracture toughness, $K_{IC}$, the shear mode fracture toughness $K_{IIC}$ is calculated from the following empirical elliptical criterion [135]

$$\left(\frac{K_I}{K_{IC}}\right)^2 + \left(\frac{K_{II}}{K_{IIC}}\right)^2 = 1 \tag{25}$$

where $K_I = F_I \sigma \sqrt{\pi a}$ and $K_{II} = F_{II} \sigma \sqrt{\pi a}$ and $F_I$ and $F_{II}$ are obtained from the empirical relations with less than 0.5% error for a centre inclined crack in a finite plate subjected to tensile stress [134].

4. Results and Discussion

4.1. Stress- Strain Curve

Fig. 13 shows the plots of stress-strain curves for random and aligned PLA fibrous networks. Both cases of untreated and doped nano-fibrous mats are simulated for each configuration. For the sake of comparison and to ensure the same density, the number of particles is kept almost the same for the aligned and random configurations.

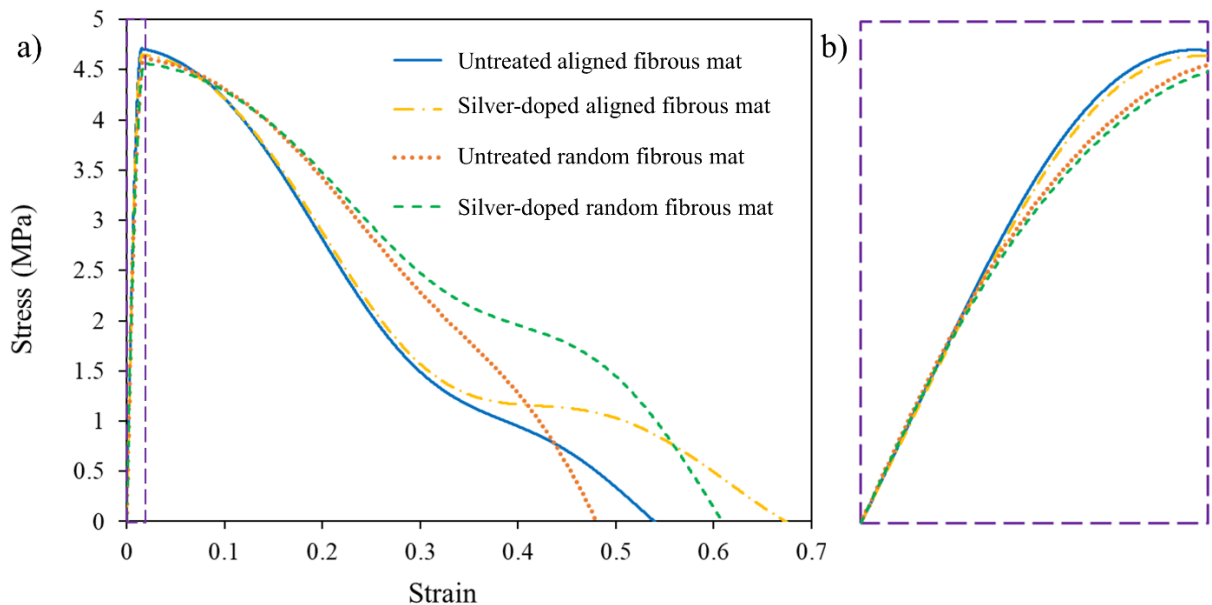

Fig. 13. a) Stress-strain plots for simulated PLA fibrous networks with different fibre orientation and doping agent, and b) Enlarged linear part for more clarification

Using the stress-strain curves, Table 2 provides a list of the Young's modulus and tensile strength obtained from simulations on fibrous networks with different fibre orientation and surface treatment. Besides, some experimental data from the literature are presented in Table 2 where a good agreement can be found with the simulation results of the current work. However, it should be noted that various factors, such as straightness and diameter of the nanofibres as well as the quality of interconnections can affect the network mechanical parameters.

To obtain $K_{IC}$, for each nanofibrous network, the maximum stress obtained from simulations of pre-cracked network with horizontal crack is used with Eq. (24). Further, the maximum stress obtained for the inclined crack is used together with Eq. (18) to obtain $K_{IIC}$ which is presented in Table 2. In Eq. (25), the value of $F_I$ and $F_{II}$ for $\frac{a}{W} = 0.4$ and $\beta = 60°$ are calculated as 0.8456 and 0.4497, respectively based on the empirical relation in [134].

| | Reference | Characterization Method | Orientation | Treatment | Young's Modulus (MPa) | Tensile Strength (MPa) | Fracture Toughness KIC (MPa√m) | Fracture Toughness KIIC (MPa√m) |
|---|---|---|---|---|---|---|---|---|
| 1 | Present Work | Hierarchical MD-PD simulation | Aligned | Untreated | 432 | 4.71 | 0.0927 | 0.0575 |
| 2 | | | Aligned | Doped with Ag Nanoparticle | 421 | 4.65 | 0.0915 | 0.0568 |
| 3 | | | Random | Untreated | 411 | 4.63 | 0.0910 | 0.0570 |
| 4 | | | Random | Doped with Ag Nanoparticle | 402 | 4.57 | 0.0898 | 0.0562 |
| 5 | Magiera et al, 2017 [136] | Experiments | Random | Blended with Gelatine | 459 | 4.1 | - | - |
| 6 | Lopresti et al, 2020 [33] | | Aligned | Doped with HA Microparticle | 400 | 8 | - | - |
| 7 | Magiera et al, 2017 [136] | | Random | Untreated | 686 | 10.6 | - | - |
| 8 | Lopresti et al, 2020 [33] | | Aligned | Untreated | 250 | 7 | - | - |

4.2. Effect of Fibre Orientation on the Stress Strain Response

From the stress-strain curve presented in Fig. 13 and the Young's modulus data reported in Table 2, the fibrous mat characterised by aligned fibres exhibits greater stiffness and Young's modulus within the elastic and linear phase. However, fracture initiates earlier and at a higher stress threshold. This shows that the aligned network exhibits more brittleness compared to the random arrangement. Aligned nanofibres gives the structure an orthotropic nature with higher stiffness in the fibre direction, while, random network resembles an isotropic but less stiff material. On the other hand, localized areas of high stress and reduced fibre-fibre interactions can accelerate the initiation of fractures in aligned networks. This observation is consistent with the experimental findings of Lopresti et al. [33] where the Young's modulus of PLA-based electrospun scaffolds were higher in aligned fibre arrangement compared to the random one. Besides, the fracture toughness in fibrous network is improved by the aligned arrangement (as shown in Table 2).

### 4.3. Crack propagation

Fig. 14, represents the crack evolution and damage patterns in the aligned pre-cracked fibrous network doped with metallic nanoparticles where the colours represent the damage value of the particles.

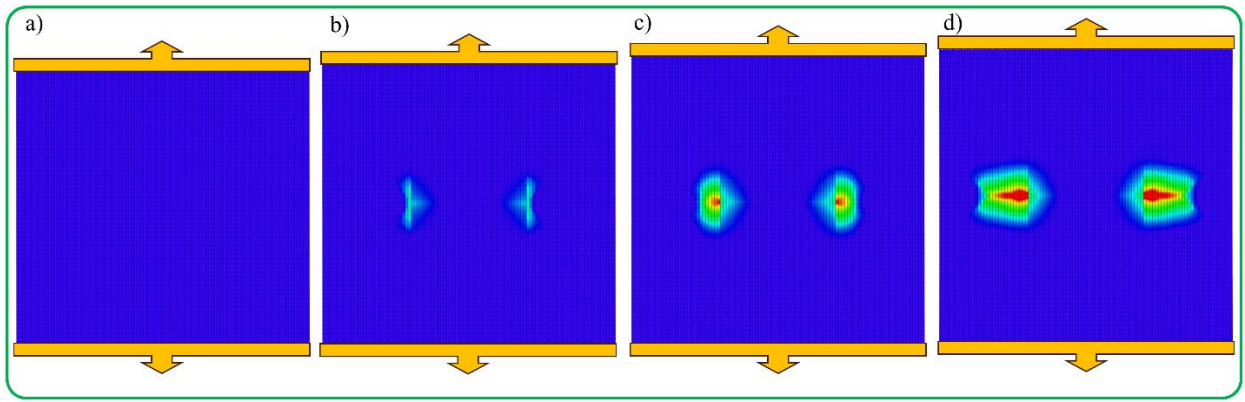

Fig. 14 Crack evolution and propagation in PLA fibrous network doped with silver nanoparticle and with aligned distribution of fibres: a) linear stress strain relation and no crack, b) ultimate stress and crack initiation, c) decreasing stress and crack propagation, and d) zero stress and sever damage

Fig. 15a- 15d shows the crack propagation and damage patterns for the fibrous network with random distribution of fibres.

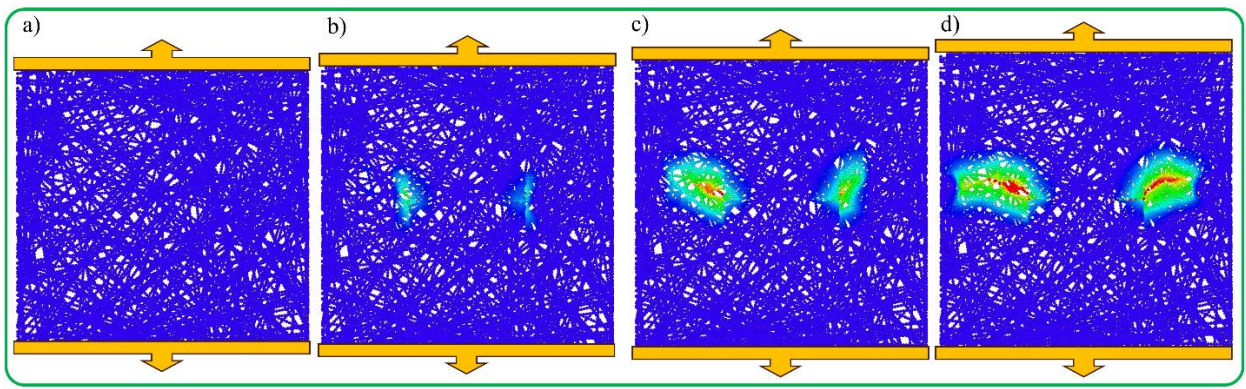

Fig. 15 Crack evolution and propagation in PLA fibrous network doped with silver nanoparticle and random distribution of fibres: a) linear stress strain relation and no crack, b) ultimate stress and crack initiation, c) decreasing stress and crack propagation, and d) zero stress and sever damage

It can be noted in Figs. 15b to 15c that unlike the aligned fibrous network, in randomly oriented fibrous network the crack does not grow in a straight direction perpendicular to the load direction, but the crack path deviates due to the irregularities in the internal structure. Also considering the stress comparison in Fig. 16, it can be concluded that although the aligned configuration has a higher stiffness before crack propagation and the stress value required for a given strain is higher, after the crack starts to propagate, the randomness of the fibres' alignment and their interconnection hinders the crack propagation in the random fibrous network, thus impeding the damage-induced stress reduction.

Fig. 16 addresses the points on the stress strain curve related to various phases of loading in Fig. 14 and Fig. 15. which demonstrates that in Fig. 14a (and 15a), no damage has occurred, and the fibrous network is in the elastic linear phase of strain and stress. The pre-crack begins to propagate (in Fig. 14b and 15a) that leads to the emergence of nonzero damage values ahead of the crack tips. This step corresponds to the peak value on the stress curve. As the crack grows (Fig. 14 c-d and 15. c-d), the number of particles with nonzero damage values increases, and the damage parameter rises in the already damaged particles. Consequently, the load (i.e., stress) carried by the plate decreases due to the reduction in plate stiffness.

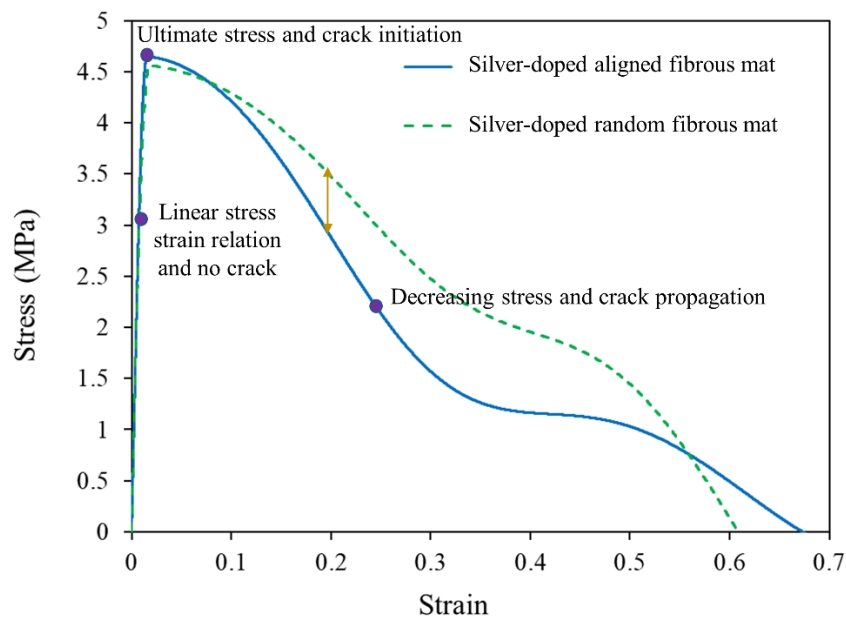

Fig. 16. The points on the stress-strain curve relating to different loading phases for silver-doped aligned and random fibrous networks demonstrating the linear elastic phase without damage, the crack initiation reflected in the stress drop, and the stress relief accompanied by crack propagation

In the case of inclined cracks in the aligned fibrous network, the angle between the crack path and the direction of the initial crack ($\theta_0$ in Fig. 17) is measured and compared to the analytical estimation in [137] for a homogenous isotropic plate. According to [137], the crack extension angle, $\theta_0$, for a homogenous isotropic plate with $\beta = 60°$ is around 45°, whilst in the simulated aligned fibrous network it is increased to 70°. The increase in the crack extension angle can be originated from the weaker load transfer in the horizontal direction compared to the vertical one in the aligned network. This is also consistent with another study [138] on the crack extension angle of the inclined crack in an orthotropic medium which confirms that the crack path tends to the stiffer material direction.

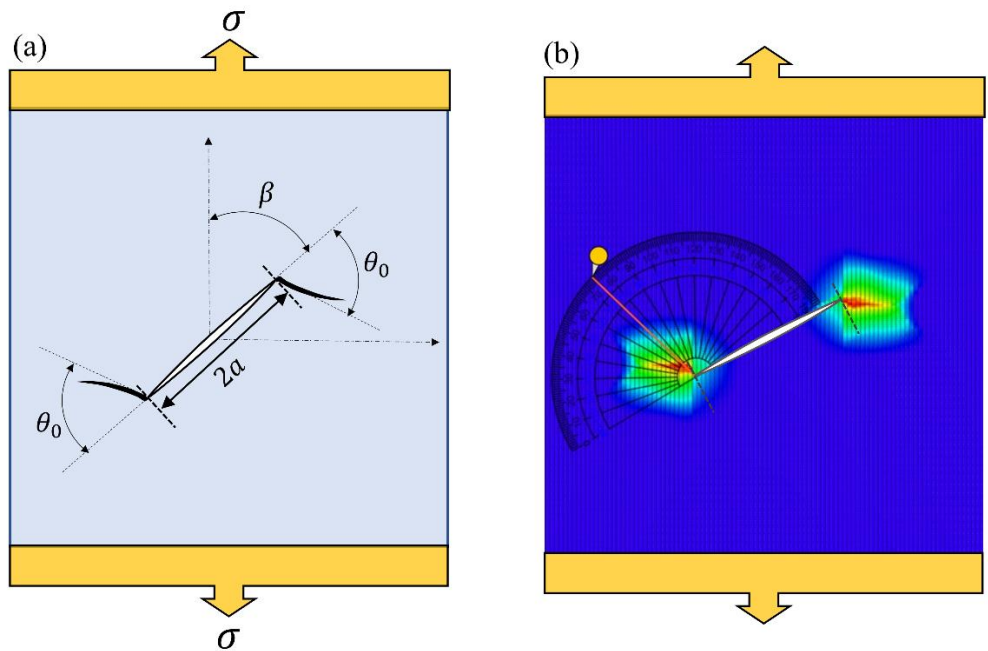

Fig. 17. The schematic crack path in a plate with an inclined crack with respect to the load direction, and b) The measured angle $\theta_0$ for the crack path in the simulated aligned fibrous network.

Note that surface effects arise near boundaries due to the imbalance of forces experienced by material points at or near the surface, which can lead to altered stress states and influence crack behaviour. This phenomenon is particularly significant in the context of peridynamics, where interactions are inherently non-local, and the absence of neighbouring material at boundaries leads to higher surface energies and stress concentrations. To avoid these boundary-induced surface effects, as seen in Fig. 14 and 15, we have limited the crack propagation analysis to regions within one peridynamic horizon ($\delta$) away from the vertical boundaries. This is because the surface effects, though pronounced near boundaries, diminish significantly within one to two horizon distances into the bulk material [139].

### 4.4. Effect of Surface Treatment with Silver Nanoparticles

Comparing the results of silver doped networks with the counterpart untreated ones reveal that doping with nanoparticles slightly reduces the Young's modulus. This is in line with the experimental observations of Cacciotti et al [19], who studied the effects of silver nanoparticles on the mechanical properties of random electrospun PLA mats.

However, the main effect of the doping particles occurs after crack initiation. In this case, they impede crack growth and increase the elongation at break. This phenomenon is also observed in the MD simulation of a single nanofibre in Section 2, whereby the nanoparticles play their effective role by maintaining the integrity of the polymer chains after crack initiation. This suggests that the proposed method can preserve the physical properties of the material during hierarchical transfer. The persistence of this post-crack initiation effect in the PD description at the microscale further justifies the suitability of employing the PMB model in this study, despite its simplicity.

### 4.5. Computational considerations and Expected Limitations

In this subsection, we present some notes about the computational considerations and limitations of the current study.

- To improve the conditioning and well-posedness of the problem, different features are considered in the current work; The implementation of the staggered scheme rather than the monolithic one simplifies the computational process and avoid dealing with sparse matrices which is generally encountered in the monolithic approach and may lead to ill-posed problems. Besides, for microscale implementation, according to several studies, in the non-local PD model, the condition number of the stiffness matrix, $\kappa(K)$, has an upper bound with the same order as $\delta^{-2}$ [140-143]. Specifically in [140], through a spectral equivalence, the authors have shown that $\kappa(K) \leq \delta^{-2}$ and there is no independence of the upper bound to the mesh size. In the current work, the horizon is determined as detailed in Section 3.2 resulting in a condition number less than 44.45 for the domain simulations. Also, consistent with [140], the condition number is almost independent to the mesh size. At the nanoscale implantation, the reliability of the simulations is improved by adopting all-atom molecular dynamics simulations instead of the coarse-grained one. Also, the

- velocity Verlet algorithm used to integrate the Newtonian equations of motions, has an error on the order of $O(\delta t^3)$ with $\delta t$ the time step. Velocity Verlet integrator conserves energy well over long simulations, making it stable and well-conditioned for MD simulations [144]. On the other hand, the choice of 0.5 femtosecond for the timestep which is 1/20 of the highest vibration frequency related to the carbon-hydrogen bond, is small enough to accurately resolve the fastest motions in the system and prevents significant numerical errors from accumulating [145].
- In the current coupling approach, the peridynamic model, although primarily focuses on tensile bond failure, can account for shear effects through the combination of bonds oriented in different directions. In the bond-based PD model, the material response to shear loading is represented by the collective behaviour of multiple bonds acting in various directions. As shear forces are applied, these bonds experience different levels of stretch or compression, which together contribute to the material's overall resistance to shear deformation. Moreover, the integration of MD simulations into our coupling approach enhances the model's ability to capture a broad range of more complex mechanical behaviours, including shear. Using atomistically derived bulk modulus and energy release rate in the PMB PD model allows the shear behaviour at the lower scale to be implicitly incorporated. The parameters derived from MD simulations reflect the combined effects of tension and strengthening of polymeric chains as well as their slippage and shear at the atomic level, which are then integrated into the PD framework and influence how the bonds behave in the PD simulation. While the bond-based PD model captures shear effects through the interaction of bonds, we acknowledge its limitations in fully modelling shear-dominated failure modes. Future extensions of this work could involve the implementation of state-based PD models or the use of more enhanced bond-based PD models.
- To provide an estimate of the computational efficiency gained through multiscale modelling, we report that simulating a single nanofiber using MD on a 64-bit server with 88 cores and 256 GB of RAM took approximately 780 hours. In contrast, simulating the same nanofiber using PD with data derived from the MD simulation required only about 7 minutes. Extending the PD simulation to a full fibrous network took approximately 326 hours. This comparison highlights that simulating a fibrous network of the dimensions studied in this work would be nearly infeasible using a fully MD approach.

- Both MD and PD simulations were performed using MPI-based parallelization, which significantly reduced computational time. However, efficient data management and communication between processors are essential to ensure proper parallel implementation and to avoid bottlenecks that could damage these time savings.
- The peridynamics model employed in this study is the prototype microelastic brittle (PMB) model, which, despite its simplicity, has proven effective in capturing fracture behaviour. However, the PMB model's reliance on a limited set of material parameters constrains its ability to fully replicate the complex stress-strain responses observed in MD simulations, which often include multiple phases of softening and hardening during deformation. More advanced models, such as state-based PD, elastoplastic [146] or viscoelastic peridynamics [147] which incorporate additional parameters, can provide a more accurate representation of the stress-strain curves, including plasticity, that are not fully addressed by the PMB model. However, these advanced models also increase computational cost and complexity, particularly in large-scale simulations.

5. Summary and Conclusion

In the present work, a hierarchical approach bridging nano to micro scale is presented to study the elastic and fracture parameters of nanofibrous network with random and aligned distribution of nanofibres. The elastic and fracture parameters of single untreated and silver doped nanofibres are determined using molecular dynamics simulation and the parameters are used to inform the micro scale described in the framework of PD. Subsequently, the effect of fibre distribution and composition on the Young's modulus, tensile strength, mode I and mode II fracture toughness is studied. In addition to study the effect of fibre distribution, the effect of surface treatment on the obtained parameters are investigated. The following conclusions can be inferred from this study:

- The surface treatment with silver nanoparticles slightly reduces the Young's modulus. This is in line with experimental observations of Cacciotti et al. [19], who studied the effects of silver nanoparticles on the mechanical properties of random electrospun PLA mats. However, the main effect of the doping particles occurs after crack initiation, whereby they impede crack growth and increase the elongation at break.
- The simulated values of Young's modulus and tensile strength for untreated and doped random fibrous networks are in agreement with the experimental data in literature.

- The dominant effect of nanoparticles on the crack growth is observed both in the MD and PD approaches. This suggests that the proposed method can preserve the physical properties of the material during hierarchical transfer.

- In contrast to aligned fibrous networks, cracks in random fibrous networks do not propagate in a straight line perpendicular to the load direction. Instead, the crack path deviates due to irregularities in the internal structure.

- The aligned fibrous network exhibits higher stiffness compared to the random fibrous network prior to crack propagation. This is consistent with experimental observations of Lopresti et al. [33]; however, after the crack starts to propagate, the randomness of the fibres' alignment in the random fibrous network and their interconnection hinder the crack propagation, thus impeding the damage-induced stress reduction.

- For an inclined crack in an aligned fibrous network, the angle between the crack path and the original crack direction exceeds the analytical estimates that apply to homogeneous isotropic plates. This result is consistent with previous research on crack extension angles in orthotropic media and indicates a preference for crack propagation along stiffer material directions. This deviation could be due to weaker load transfer in the horizontal direction compared to the vertical direction within the aligned network

**Funding**
This work has been supported by Italian Ministry of University and Research [no. 1062 of 10/08/2021, the endowment of the PON "Research and Innovation" 2014-2020]. The support of research project PNRR - CN1- Spoke6 – CUP B83C22002940006 and PRIN 2022-2022YLNJRY (grant number: J53D23002500006) Funded by the European Union – Next Generation EU is also acknowledged.

The Authors wish to acknowledge Dr Greta Ongaro for her support in understanding the basic concepts of Peridynamics and in writing Section 3.1